# Controlled formation of high-mobility shallow electron gases in SrTiO$_3$ single crystal

Jung-Won Chang[1], Joon Sung Lee[1]*, Tae Ho Lee[1], Jinhee Kim[2], and Yong-Joo Doh[1]*

[1]*Department of Applied Physics, Korea University Sejong Campus, Sejong 339-700, Republic of Korea*

[2]*Korea Research Institute of Standards and Science, Daejeon 305-600, Republic of Korea*

E-mail: simmian@korea.ac.kr, yjdoh@korea.ac.kr

We report controlled formation of sub-100 nm-thin electron channels in SrTiO$_3$ by doping with oxygen vacancies induced by Ar$^+$-ion irradiation. The conducting channels exhibit a consistent high electron mobility (~15,000 cm$^2$V$^{-1}$s$^{-1}$), which enables clear observation of magnetic quantum oscillations, and gate-tunable linear magnetoresistance. Near the onset of electrical conduction, the metal-insulator transition is induced by the mobility suppression. With the high electron mobility and the ease of controlled channel formation, this ion-irradiation doping method may provide an excellent basis for developing oxide electronics.

SrTiO$_3$ (STO) has the potential to become a key ingredient in oxide electronics[1] owing to its intriguing electronic properties in addition to its versatility as a perovskite substrate for oxide heterostructures.[2,3] The recent advancement in studies of the low-dimensional electronic system formed at the heterointerface between strontium titanate (STO) and lanthanum aluminate (LaAlO$_3$; LAO)[4] has led to immense development and success in both understanding fundamental physics and developing new oxide electronic devices. When pure and stoichiometric, STO is an insulator with a large band gap of 3.2 eV. However, it can be made into an $n$-type conductor by substitutional doping with cations[5,6] such as Nb and La or by introduction of oxygen vacancies.[7] Especially, Ar$^+$-ion irradiation on the crystal surface[8-13] introduces oxygen vacancies and enables us to pattern conducting channels selectively on STO substrates using photolithography. Most previous works on irradiation-doped STO have dealt with samples with extensive irradiation.[8-11] In this work, we focus on very shallow irradiation conditions that cover the onset and subsequent proportional increase of conductance. We demonstrate that the sheet carrier density of the conduction channel is directly controlled by the irradiation time while maintaining high electron mobility, which enables pronounced quantum oscillations below 10 K. Near the onset of conductivity, gate-tunable linear magnetoresistance (MR) is observed, the ratio of which reaches 700% at 9 T. Furthermore, sub-100 nm-thin samples near the conductivity onset exhibit metal–insulator transition when a small gate voltage $V_g$ is applied at low temperatures. Our work provides a simple and reliable way to form a very thin high-mobility conduction channel beneath the surface of STO, which is applicable to studies on low-dimensional quantum transport phenomena[14] combined with superconductivity[6,15] and ferromagnetism[16] in strongly correlated metal oxide systems.

Commercial single crystals of STO[001] with a thickness of 0.5 mm were used for our experiments. Au/Ti (10/50 nm) layer was sputtered on the backside of the STO substrates for the gate contact. The bridge-type Hall bar geometry (1000 μm × 100 μm) shown in Fig. 1(a) was defined on our samples using photolithography. Then, the STO samples were irradiated by Ar$^+$ ions accelerated by a beam voltage of 450 V with varying the irradiation time $t_{irr}$ from 20 s to 100 s. The samples were labeled according to $t_{irr}$; for instance, the sample corresponding to $t_{irr} = 50$ s was labeled as S50. The ion beam had a current density of 0.7 mA/cm$^2$ and an incident angle of 45°. The areas protected by the photoresist mask were insulating, and the STO samples irradiated without the Hall bar-shaped opening showed an electrical resistance over 1 GΩ at room temperature. Ohmic contacts were made on the large contact pads of the irradiated STO by using Al wire wedge bonding. The transport properties of the samples were measured in a physical property measurement system from Quantum Design with a rotating sample holder. The temperature- and gate-dependent transport characteristics were measured by a four-probe dc measurement scheme with a bias voltage of 1 mV, and the magnetotransport properties were measured by an ac lock-in technique with a bias current of 100 nA.

Figure 1(b) shows the temperature dependence of the sheet resistance $R_s$, the sheet carrier density $n_s$, and the Hall mobility $\mu_H$ for samples S20–S100. Except for S20, which becomes insulating below $T = 200$ K, all the samples exhibit apparently similar metallic characteristics: $R_s$ decreases with decreasing $T$, $n_s$ does not freeze out down to 2 K, and $\mu_H$ reaches ~15,000 cm$^2$V$^{-1}$s$^{-1}$ for $T < 10$ K. It should be emphasized that this mobility value, although not yet optimized, is comparable to the highest one obtained from the La-doped STO grown by molecular beam epitaxy.[5] From a practical point of view, it is remarkable that the range of high mobility typically observed from optimally doped bulk STO[7] can be achieved by such a simple fabrication method.

The residual resistivity ratio RRR ($\equiv R_s(300$ K$)/R_s(2$ K$)$) is about 2000, which is comparable to the values obtained from high-mobility bulk STO samples lightly doped with oxygen vacancies or Nb.[7,17] Figure 1(c) shows the dependence of $R_s$ and $n_s$ on $t_{irr}$. It is observed that $n_s$ increases linearly with $t_{irr}$ above a threshold time of about 20 s, while $\mu_H$ is not affected by the irradiation time at least up to $t_{irr} = 100$ s. Gate-voltage dependence of the sheet conductance, $G_s = 1/R_s$, exhibits a sub-threshold swing of ~20 mV/dec with a threshold voltage of −3 V for S27. Our results imply that the low-dose Ar$^+$-ion irradiation technique is a very convenient way to pattern high-mobility electron channels in STO with excellent gate modulation, in addition to the controllability of $n_s$ and $R_s$ via $t_{irr}$. Our work is distinct from previous works[8-11] reporting relatively low-mobility channels formed with large irradiation dosage[8,9,12] or much higher ion energy.[10]

In an external magnetic field $B$, the doped STO samples exhibit very distinct $t_{irr}$-dependent MR curves at low temperature. Figure 2(a) displays MR curves obtained from S50 with a magnetic field applied perpendicular to the sample plane. The parabolic MR curves for $T = 30$ K and 50 K merge into a common curve with the relation $\Delta R(B)/R(0) \sim f(\mu_{MR}B)$, where $\mu_{MR}$ is the MR mobility and $\Delta R(B) = R(B) - R(0)$, indicating that Kohler's rule[18] holds at higher temperatures (see Fig. S1 in Supplementary data). Similar behavior is observed for S23 in Fig. 2(b). At low temperatures below 10 K, however, the MR for S50 becomes saturated incompletely with increasing $B$, whereas the MR for S23 keeps increasing linearly. The MR for S100 is similar to that for S50, and other samples with $t_{irr} \leq 30$ s behave similarly to S23 (see Fig. S2). The incompletely saturating MR for S50 can be separated into two components: a saturating quadratic MR of $\Delta R(B)/R(0) \sim B^2/[1+(\mu_{MR}B)^2]$, as expected for conductors with closed Fermi surfaces[19], and a linear MR, which will be discussed later. Fitting of the MR curves by combining these two effects yields $\mu_{MR}$ values very close to the $\mu_H$ values determined by Hall measurements (see Fig. S3), supporting the validity of the MR formula as a description of the MR observed in the highly doped STO samples.

Occurrence of large linear MR is usually explained in terms of quantum magnetoresistance[20,21] or mobility disorder.[22] Abrikosov's quantum magnetoresistance, however, requires either only the lowest Landau level to be partially occupied[20] or an effective linear energy dispersion relation near the Fermi level,[21] neither of which is met by our samples (see the Supplementary data). Therefore, another explanation by disorder- or inhomogeneity-induced linear MR would be more plausible. It is known that electrical current flowing under an external magnetic field through a sample with conductivity inhomogeneity can create locally inhomogeneous Hall fields, and the Hall fields produce local distortions in the current path, eventually resulting in inhomogeneity-induced geometrical MR. Also, Herring[23] has shown that the current distortions caused by inhomogeneity extend in the direction of the applied $B$ to increase the geometrical MR, with the range of the distortion nearly proportional to the field strength. The inhomogeneity in the doped STO is presumed to occur near the interface between the top layer damaged by the $Ar^+$-ion irradiation[10] and the less disturbed conduction channel beneath it. In such a case, the normal field MR should be larger than the MR under $B$ parallel to the sample plane, because the above-mentioned current path distortion would be propagated only along the direction of $B$, thus causing only negligible geometrical MR under plane-parallel $B$. This is consistent with our observation; the MR of S23 under normal $B$ (Fig. 2(b)) is about 40 times larger than that under longitudinal $B$ (Fig. 2(d)). Since the region farther from the inhomogeneity is less affected by the current path distortion, the normal field MR is expected to be enhanced for thinner channels. The normal field MR of S23 is about seven times larger than that of S50, and the MR ratio increases up to 700 % at $B = 9$ T under a negative $V_g$ in Fig. 2(c), which conforms to the above inference.

It should be noted that the longitudinal field MR in Fig. 2(d) contains multiple kink structures superimposed on the linear MR at low temperatures. Subtracting the background signal reveals the resistance oscillations as a function of an inverse magnetic field, as depicted in Figs. 3(a)–3(c). These oscillations are attributed to the Shubnikov–de Haas oscillations (SdHO), which are quantum oscillations in electrical conductivity modulated by the alignment between the Fermi level and the quantized Landau levels. We have analyzed the SdHO using the following simplified formula originating from Roth and Argyres:[24]

$$\frac{\Delta R}{R(0)} \sim \sqrt{\frac{\hbar \omega_c}{2E_F}} \frac{2\pi^2 k_B T / \hbar \omega_c}{\sinh(2\pi^2 k_B T / \hbar \omega_c)} \exp\left(-\frac{2\pi^2 k_B T_D}{\hbar \omega_c}\right) \cos\left(\frac{2\pi E_F}{\hbar \omega_c} - \delta\right), \qquad (1)$$

where $\omega_c = eB/m^*$ is the cyclotron frequency, $E_F$ the Fermi energy, $\delta$ the phase factor for the oscillations, $k_B$ the Boltzmann constant and $\hbar$ the reduced Planck constant, $h/2\pi$. $T_D$ is the Dingle temperature which describes the reduction of SdHO caused by scattering-induced Landau level broadening in terms of temperature. We assumed an isotropic and parabolic dispersion relation, and omitted higher-order oscillation terms as well as the Zeeman splitting factor for simplicity (see the Supplementary data for details). Figure 3(b) shows that the magnitude of the SdHO for S23 under in-plane perpendicular $B$ is 2.5 times larger than that under longitudinal $B$, which is exactly as predicted by the theory.[24] We note that SdHO from MR curves under normal $B$ could not be extracted reliably, partly owing to the relatively large and complicated MR background, and partly due to the influence of the inhomogeneity which can dominate the MR under normal $B$, as is discussed above.

The SdHO period in the inverse magnetic field, $\Delta(1/B)$, is related to the carrier density, $n_{SdH}$, via the relation of $n_{SdH} = 1/(3\pi^2)[2e/\hbar \cdot (\Delta(1/B))^{-1}]^{3/2}$ with the assumption of a single spherical Fermi surface[17] with spin degeneracy. The observed period of $\Delta(1/B) = 0.075$ T$^{-1}$ for S50 in Fig. 3(a) corresponds to $n_{SdH} = 2.8 \times 10^{17}$ cm$^{-3}$. A comparison between $n_{SdH}$ and $n_s$ leads to a channel thickness of 3.3 μm. In a similar way, the channel thickness of S23 is estimated to be 0.36 μm using $\Delta(1/B) = 0.058$ T$^{-1}$ in Fig. 3(b). However, previous works on vertically confined conduction channels in STO have unanimously reported large discrepancy between $n_{SdH}$ and the Hall carrier density $n_H$ calculated from $n_s$ and the channel thickness.[5,6,25,26] Existence of a non-oscillatory heavy electron band has been pointed out as the cause of the discrepancy,[6] and the ratio of $n_H/n_{SdH}$ has been reported to be 3.5–4 for three-dimensional channels.[5,6] If $n_H/n_{SdH}$ of 4 is assumed for S23, its channel thickness is calculated to be ~90 nm. Also, it is noted that the SdHO for S23 is strongly suppressed for $B$ < 5 T, in contrast to that for S50. This indicates that the quantum cyclotron radius $r_c = [(2n+1)\hbar/eB]^{1/2}$ becomes comparable to half of the channel thickness at $B = 5$ T. Using a Landau level index $n = 3$ and $r_c = 31$ nm, the channel thickness of S23 is estimated to be slightly larger than 62 nm, which is consistent with the prior argument. This value is also similar to the crossover channel thickness to the two-dimensional limit reported by Kim et

al.[6]) Our method can produce sub-100 nm-thin channels in STO, of which the shallowness approaches the two-dimensional limit, although the channel itself remains three-dimensional.

From the temperature dependence of the SdHO in Fig. 3(a), the effective mass $m^*$ is determined to be ~0.16$m_e$, where $m_e$ is the bare electron mass. This value is much smaller than the range of effective mass (1.1−1.6)$m_e$ previously reported for the light carrier in STO[6]) and the value (0.5−0.6)$m_e$ for the two-dimensional electrons formed at the STO surface discovered by a recent angle-resolved photoemission spectroscopy study.[27]) We suspect that the small-valuedness of $m^*$ results from changes in the carrier density depth profile caused by the strongly temperature- and electric field-dependent $\varepsilon$ of STO at low $T$[28]) (see the Supplementary data). The gate dependence of SdHO for S50 at $T = 2$ K shown in Fig. 3(c) yields $T_D$ varying from 3.5 K to 6.0 K as $V_g$ varies from +100 V to −100 V. The quantum scattering time $\tau_q = \hbar/(2\pi k_B T_D)$ is found to be roughly proportional to the transport scattering time $\tau_{tr} = \mu_H m^*/e$ (Fig. 3(d)). Considering that $\tau_{tr}$ is generally larger than $\tau_q$ depending on the dominant scattering process,[29]) our value of $\tau_q$ obtained from the fitting is in reasonable agreement with $\tau_{tr}$.

While S50 and S100 remain electrically conducting for $V_g \geq -100$ V, other metallic samples with shorter $t_{irr}$ show an insulating transition at low $T$ under small negative $V_g$. The gate modulation of $R_s(T)$ for S25 in Fig. 4(a) shows that $R_s$ starts to diverge below 10 K near $V_g = -1.8$ V. The Hall measurement results in Fig. 4(b) reveal that the decrease in conductance near the insulating transition is dominated by the steep decrease in $\mu_H$ rather than the gradual change of $n_s$ by $V_g$. We thus infer that the insulating transition is mainly caused by the enhanced inhomogeneity suppressing the overall electron transport. As previously discussed, this inhomogeneity is also responsible for the large linear MR under normal fields.

In summary, we have studied the electrical transport properties of the shallow electron channel in STO selectively formed by a low-energy Ar$^+$ beam irradiated for short durations above a threshold time for electrical conduction. The sheet carrier density is easily controlled by the irradiation time, and the doped STO exhibits a high electron mobility of $\mu_H \sim 15{,}000$ cm$^2$V$^{-1}$s$^{-1}$ and excellent gate-controllability. The magnetic quantum oscillations are analyzed to provide further information on the electron channel, and the channel thickness has been found to be able to reach sub-100 nm regime, which is close to the two-dimensional crossover thickness. The near-interface inhomogeneity causes large linear MR and gate-induced metal–insulator transition in the samples with shorter $t_{irr}$. Consistent high electron mobility combined with the ease of channel formation proves this ion irradiation doping method to be an excellent basis for research on oxide electronics, providing scalability in the channel size down to the sub-micrometer range via conventional lithographic measures.

**Acknowledgments**

This work was supported by a Korea University Grant. We thank Soon-Gul Lee for allowing the use of Ar-ion etching system and Minu Kim and Young Jun Chang for useful discussions.

# References

## Figure Captions

**Fig. 1.** Electrical transport properties of the conducting STO samples made by $Ar^+$-ion irradiation. (a) Schematic structure of the samples. (b) Sheet resistance $R_s$, sheet carrier density $n_s$, and carrier mobility $\mu_H$ measured as functions of $T$. $n_s$ and $\mu_H$ are obtained from Hall measurements. Inset: $V_g$ dependence of sheet conductance, $G_s = 1/R_s$, for three different samples. (c) Dependence of $R_s$ and $n_s$ on the irradiation time measured at $T = 2$ K. Results of a linear and a reciprocal fitting are shown as the solid line and the dotted curve, respectively.

**Fig. 2.** Magnetoresistance curves obtained from (a) S50 and (b) S23 under $B$ applied normal to the sample plane. (c) $V_g$ dependence of the MR for S23 measured at $T = 2$ K. (d) MR for S50 and S23 under $B$ applied parallel to the current. Peaks in the superimposed magnetic oscillations are indicated by arrows. The MR curve for S23 is offset by $-6\%$.

**Fig. 3.** Magnetic quantum oscillations extracted from the MR curves by a polynomial background subtraction (a) for S50 under longitudinal $B$, (b) for S23 under longitudinal $B$ and in-plane perpendicular $B$, and (c) for S50 with $V_g$ from $+100$ V to $-100$ V with a step size of 20 V ($T = 2$ K). Landau level index $n$ is assigned to the oscillation peaks marked with arrows in (b). The solid curves are results from fitting using Equation (1). (d) The quantum scattering time $\tau_q$ obtained from the fitting of (c), compared with the transport scattering time $\tau_{tr}$ from Hall measurements.

**Fig. 4.** (a) $R_s$–$T$ data for S25 with $V_g$ from $-2.0$ V to $+2.0$ V. (b) Dependence of $n_s$ and $\mu_H$ on $V_g$ from Hall measurements on S25.

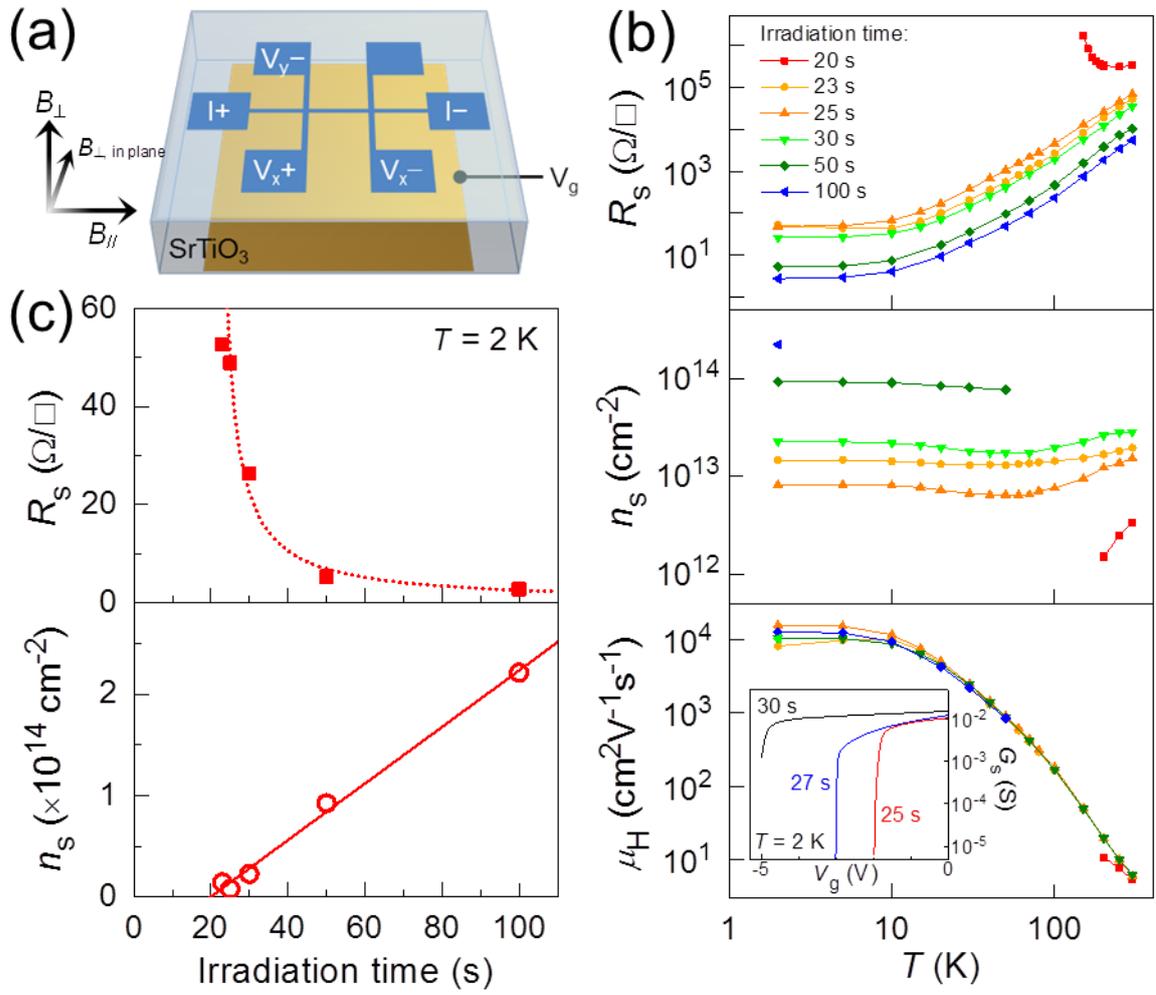

Fig. 1.

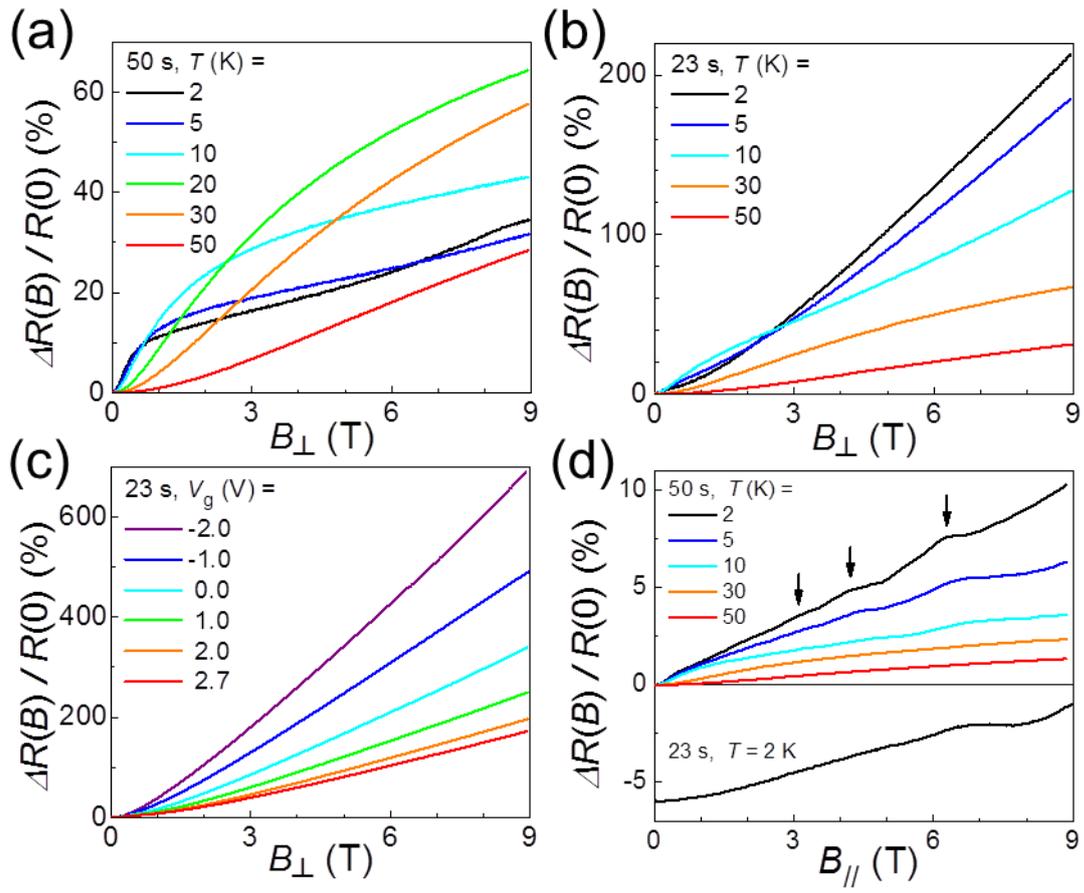

Fig. 2.

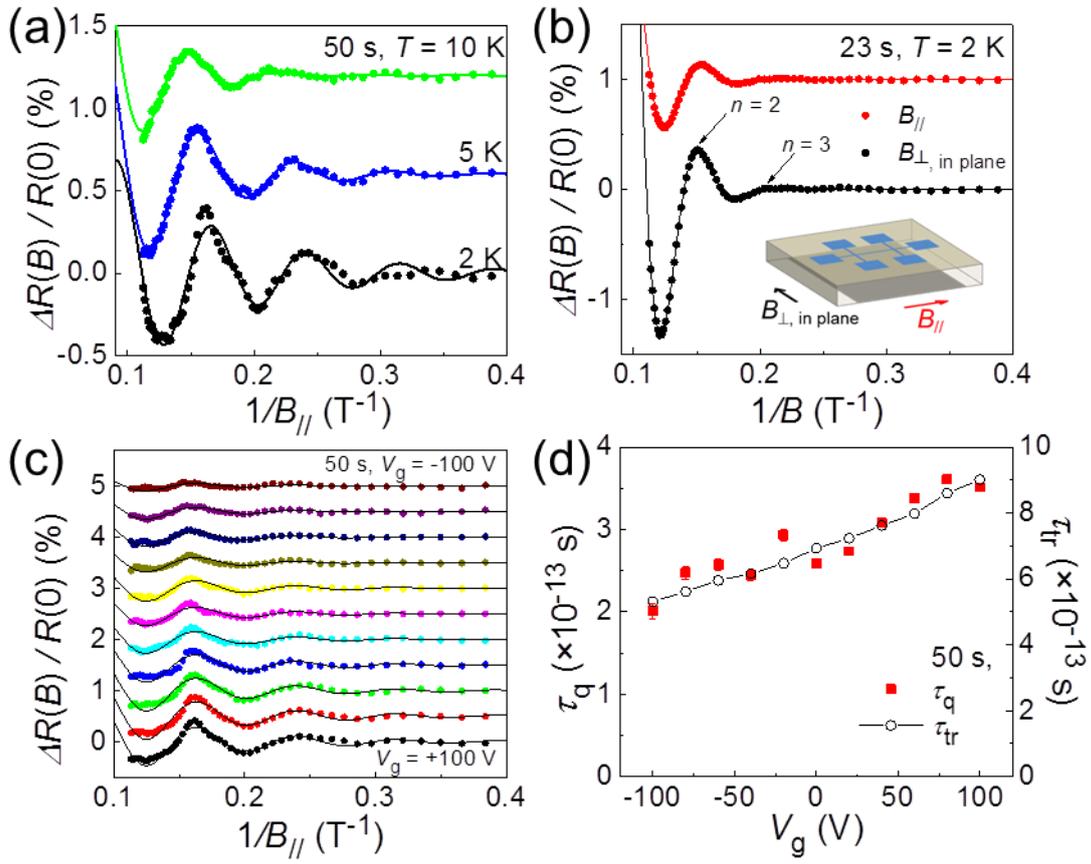

Fig. 3.

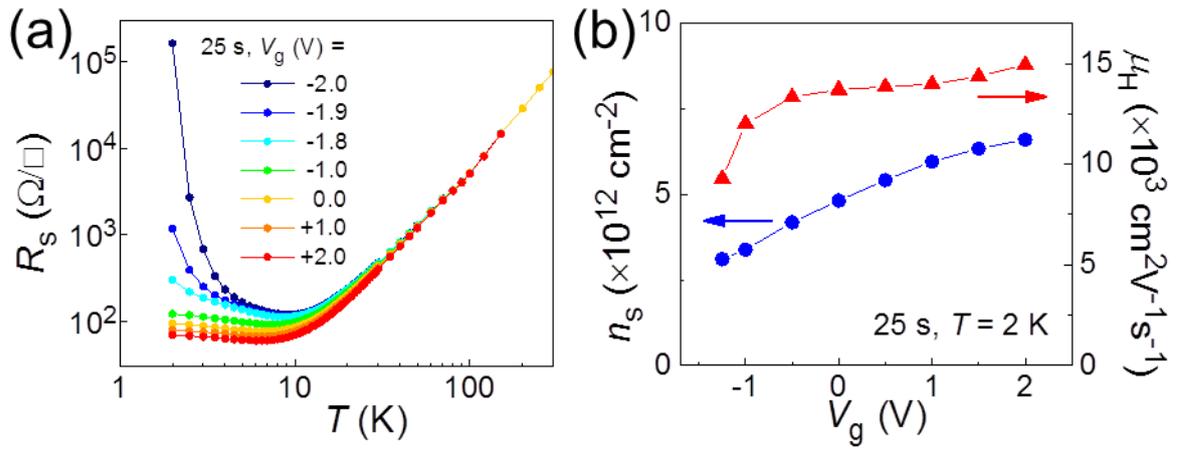

Fig. 4.